\documentclass[aps,pra,amsmath,amssymb,floatfix,twocolumn, amsmath, superscriptaddress, twocolumn]{revtex4}
\usepackage{multirow}
\usepackage{bbold}
\usepackage{subfigure}
\usepackage{color}
\usepackage{mathrsfs}
\usepackage{hyperref}
\usepackage{comment}
\usepackage{multirow}
\usepackage{amssymb}
\usepackage{pifont}

\usepackage{amsfonts, relsize, color}
\usepackage{graphics}
\usepackage{graphicx}
\usepackage{subfigure}
\usepackage{hyperref}
\usepackage{color}

\newcommand{\cmark}{\ding{51}}%
\newcommand{\xmark}{\ding{55}}%

\begin{document}
\title{Relativistic non-Fermi liquid from interacting birefringent fermions: A robust superuniversality}

\author{Bitan Roy}
\affiliation{Department of Physics, Lehigh University, Bethlehem, Pennsylvania 18015, USA}

\author{Vladimir Juri\v ci\' c}
\affiliation{Nordita, KTH Royal Institute of Technology and Stockholm University, Roslagstullsbacken 23,  10691 Stockholm,  Sweden}

\date{\today}

\begin{abstract}
We address the emergent quantum critical phenomena for (pseudo)spin-3/2 birefringent fermions, featuring two effective Fermi velocities, when they reside close to itinerant Mott transitions realized through spontaneous symmetry breaking and triggered by strong local or Hubbardlike repulsive interactions. Irrespective of the nature of the mass orderings that produce fully gapped quasiparticle spectra in the ordered phase, which otherwise can be grouped into three classes, the system always possesses a \emph{unique} terminal velocity near the corresponding quantum critical point. The associated critical regime accommodates a relativistic non-Fermi liquid of strongly coupled collective bosonic and spin-1/2 Dirac excitations with vanishing weight of the quasiparticle pole. These conclusions are also operative near superconducting critical points. Therefore, relativistic non-Fermi liquid possibly constitutes a robust superuniversal description for the entire family of strongly correlated arbitrary half-integer spin Dirac materials.     
\end{abstract}

\maketitle

\emph{Introduction}.~Notion of universality commonly incurs in the close proximity to classical and quantum continuous phase transitions, manifesting in power-law scaling of physical observables that only depends on the symmetry and dimensionality of the system~\cite{chaikin-lubensky, zinn-justin, sachdev}. A much richer, but rather sparsely encountered phenomenon is the \emph{superuniversality}~\cite{superuniversality-1, superuniversality-2, superuniversality-3, superuniversality-4, superuniversality-6, roy-kennett-yang-juricic}, extending the jurisdiction of a universality class beyond the burden of symmetry. Confluence of these two concepts can be witnessed across the quantum phase transitions (QPTs) between two topologically distinct insulators, for example. As such, the universality class of the QPT between a strong $Z_2$ topological and trivial insulators is set by \emph{odd} number (respectively, one and three in weakly and strongly correlated Bi$_2$Se$_3$ and SmB$_6$) of four-component massless charged Dirac fermions. Furthermore, massless Dirac fermions (charged or neutral) constitute a superuniversal description of QPTs between any two topologically distinct insulators belonging to any ten-fold symmetry class~\cite{topology-1, topology-2, topology-3, topology-4}.

\begin{figure*}[t!]
\includegraphics[width=0.325\linewidth]{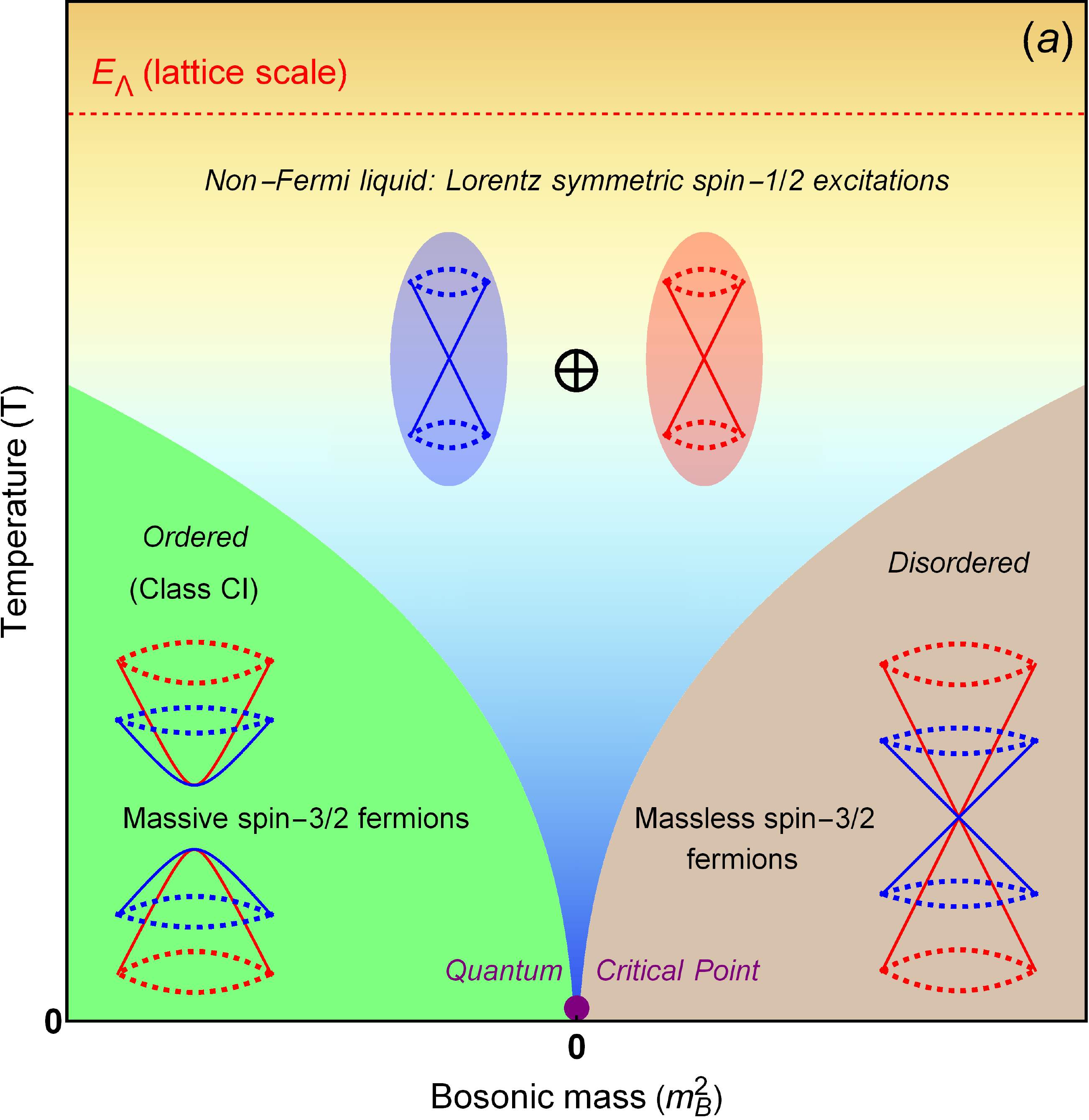}
\includegraphics[width=0.325\linewidth]{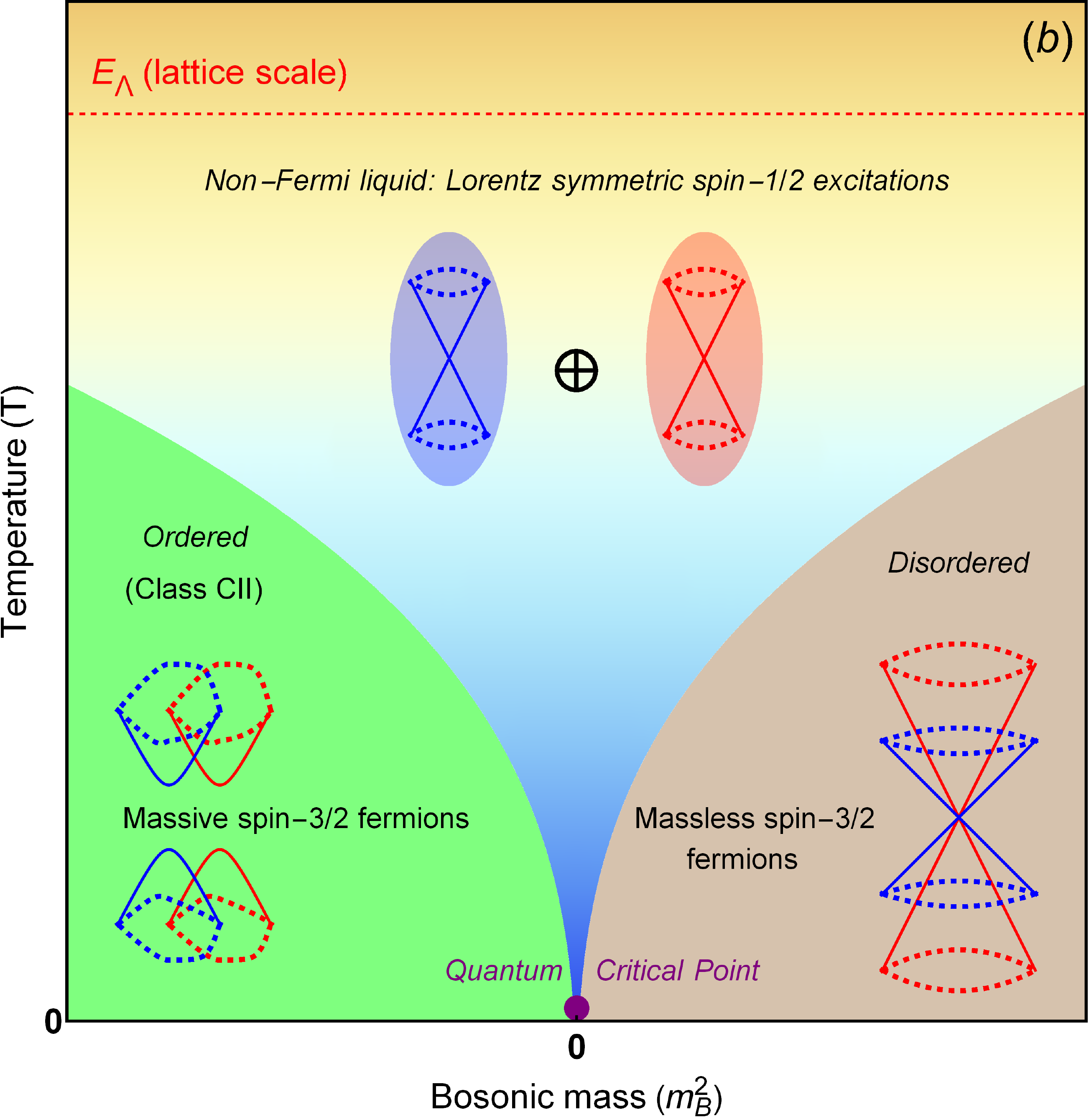}
\includegraphics[width=0.325\linewidth]{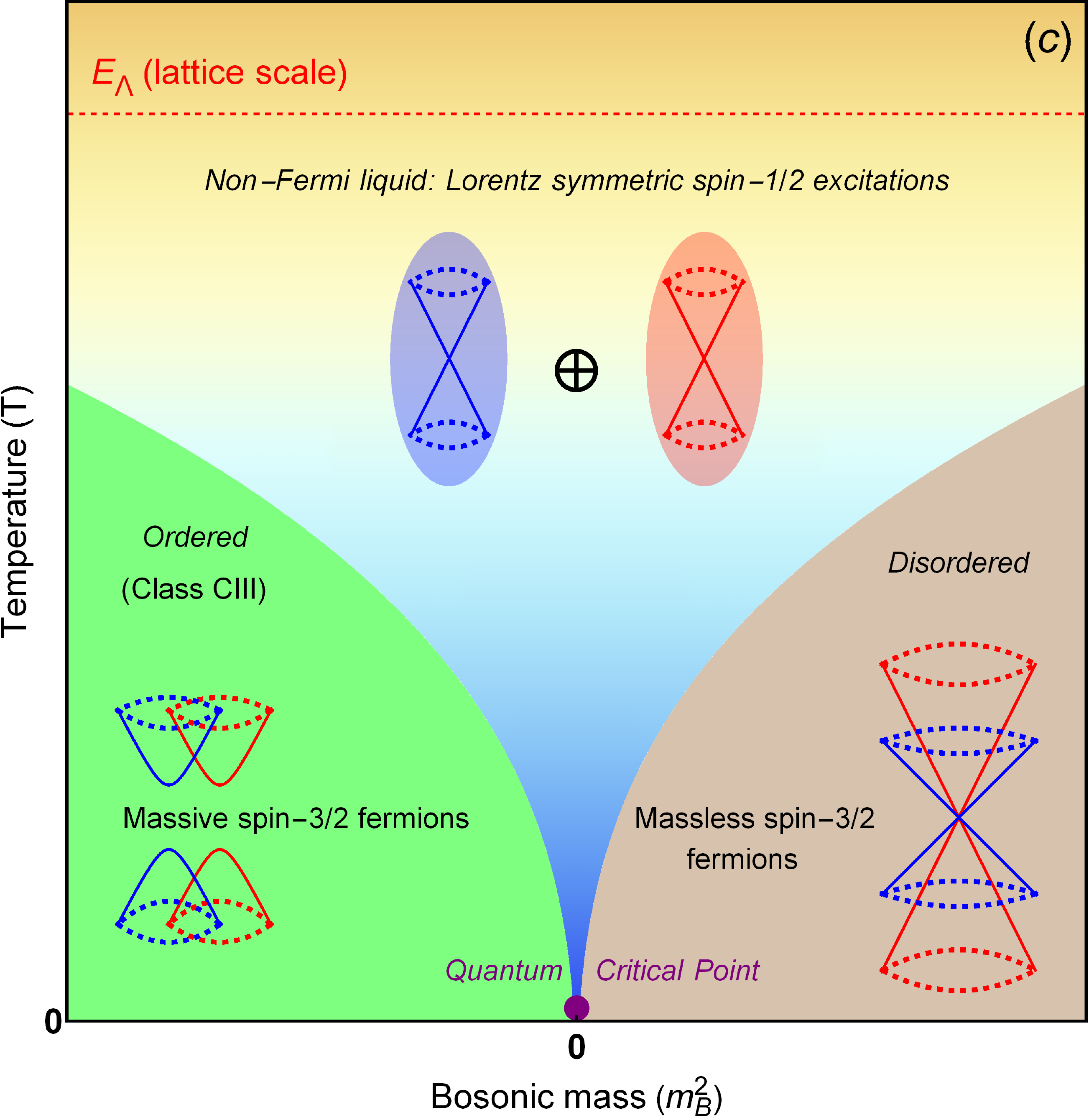}
\caption{A schematic representation of an emergent, but \emph{robust} superuniversality at the brink of Mott insulation among strongly interacting spin-3/2 birefringent fermions, with the mass ordering belonging to class (a) CI, (b) CII, and (c) CIII, see Table~\ref{table:symmetry}. Lattice realizations of mass orders from each of the classes are shown in Fig.~\ref{Fig:squarelattice}. Even though the quasiparticle spectra inside three ordered phases (green region) are distinct, at the Mott-Yukawa critical points (purple dots) and inside the quantum critical regime (thermal shaded region) the system is described by two copies of massless spin-1/2 Dirac fermions, strongly coupled with fluctuating bosonic orderparameter excitations (without any long range ordering), generically giving rise to quasiparticleless relativistic non-Fermi liquid (NFL). The signatures of the NFL, for example in optical conductivity~\cite{juricic-roy-prl}, can be observed up to a nonuniversal lattice scale $E_\Lambda$ (red dashed line). The bosonic mass ($m^2_B$) is the tuning parameter for the transitions.  
}~\label{Fig:superuniversality}
\end{figure*}

Here we show that when linearly dispersing, strongly interacting birefringent (pseudo)spin-3/2 fermions reside at the brink of itinerant Mott insulation through spontaneous symmetry breaking, the associated quantum critical point (QCP) supports a relativistic non-Fermi liquid, devoid of sharp quasiparticle excitations (due to finite anomalous dimensions for fermionic and bosonic fields), of massless (pseudo)spin-1/2 Dirac fermions, coupled with gapless collective bosonic excitations, see Fig.~\ref{Fig:superuniversality}. These outcomes do not depend on the nature of the mass orderings (otherwise, producing fully gapped quasiparticle spectra in the ordered phase, thus yielding maximal gain of condensation energy at $T=0$) and are expected to hold for any higher half-integer spin Dirac fermions, displaying multifringence of Fermi velocities. Hence, generalizing our explicit findings on strongly interacting spin-3/2 Dirac systems, we conjecture that the spin-1/2 fermion-boson coupled relativistic quantum critical state possibly constitutes a \emph{superuniversal} description for the entire family of strongly interacting multifringent Dirac fermions, which can manifest in the universal suppression of the optical conductivity~\cite{juricic-roy-prl}, for example.

\begin{figure}[t!]
\includegraphics[width=0.425\linewidth]{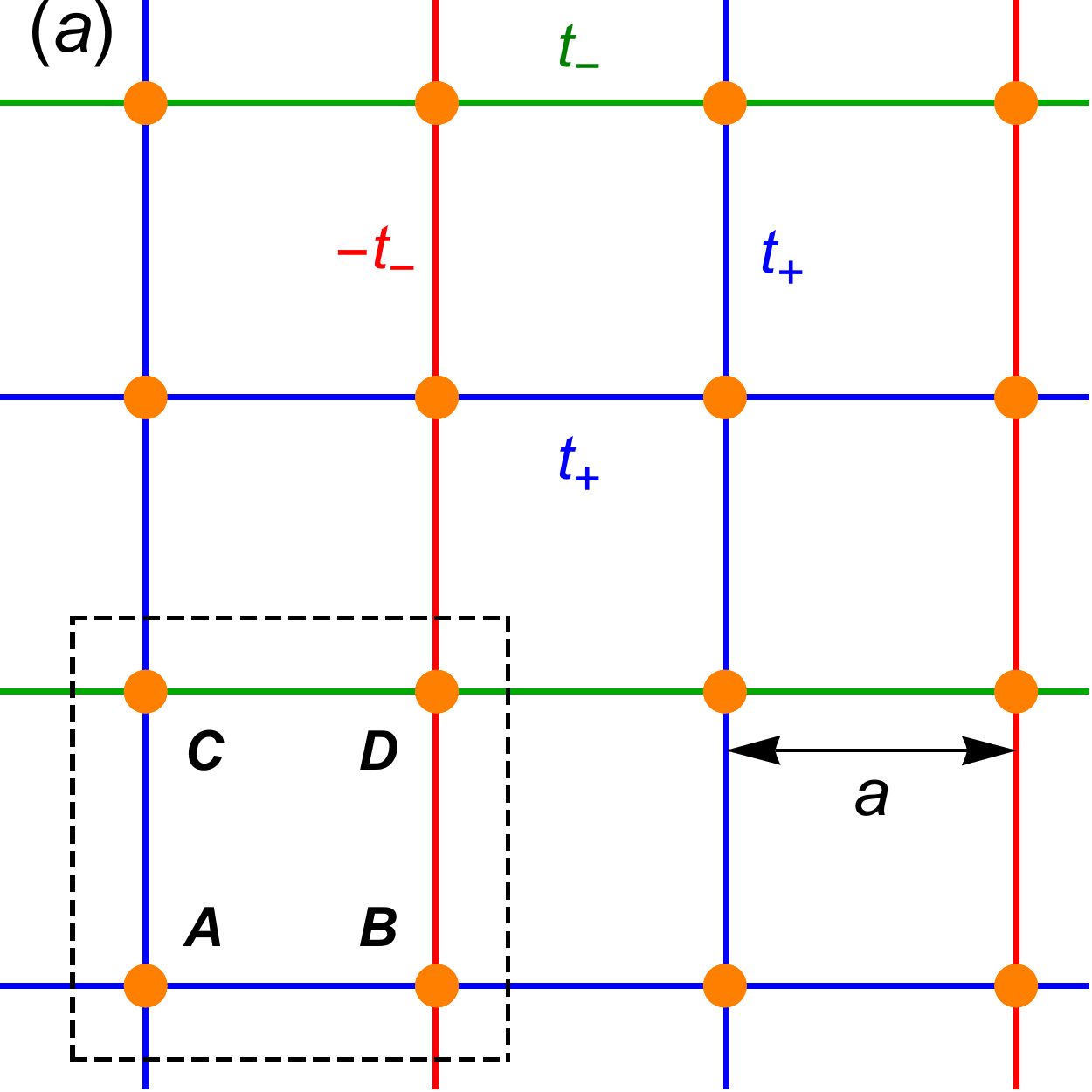}
\includegraphics[width=0.425\linewidth]{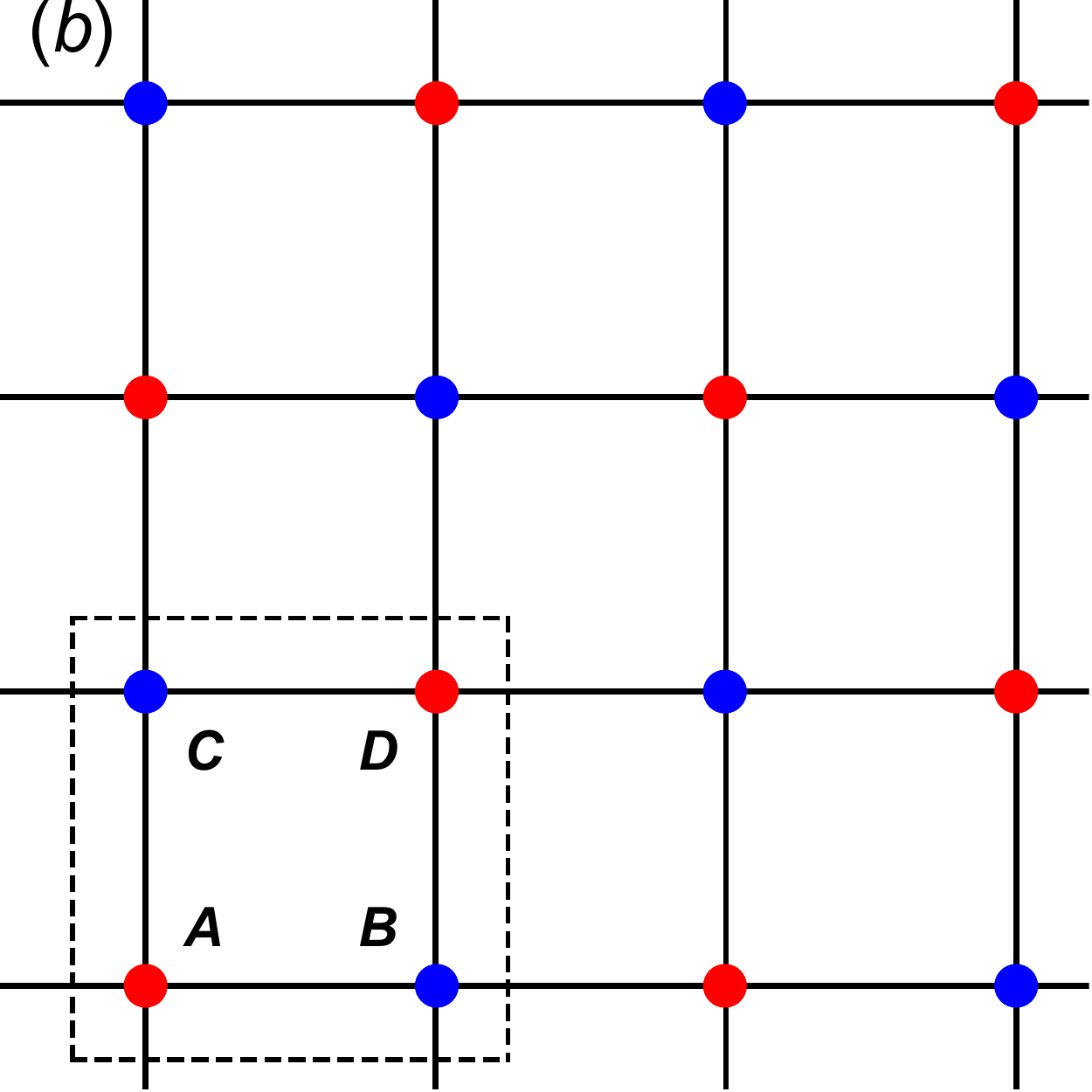}
\includegraphics[width=0.425\linewidth]{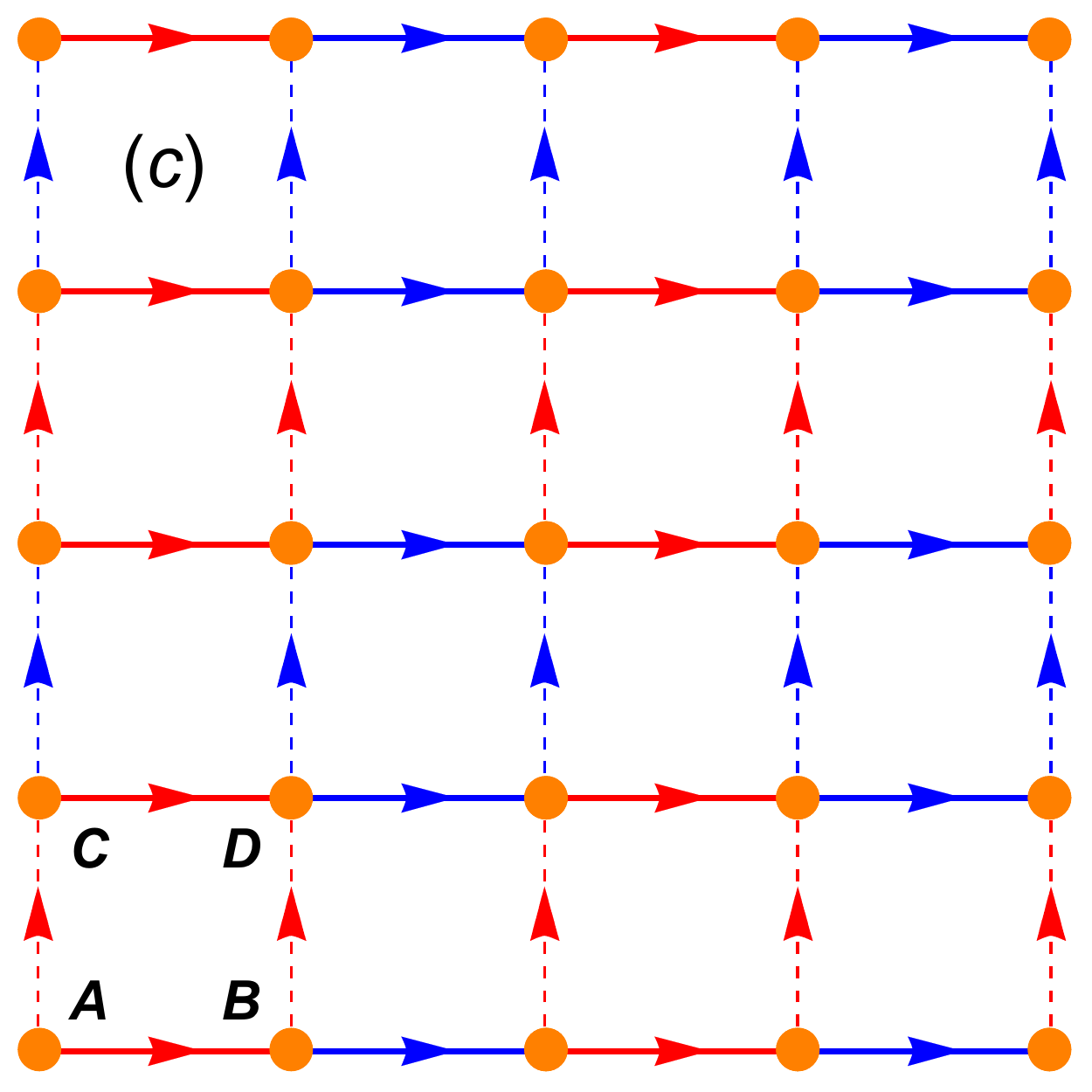}
\includegraphics[width=0.425\linewidth]{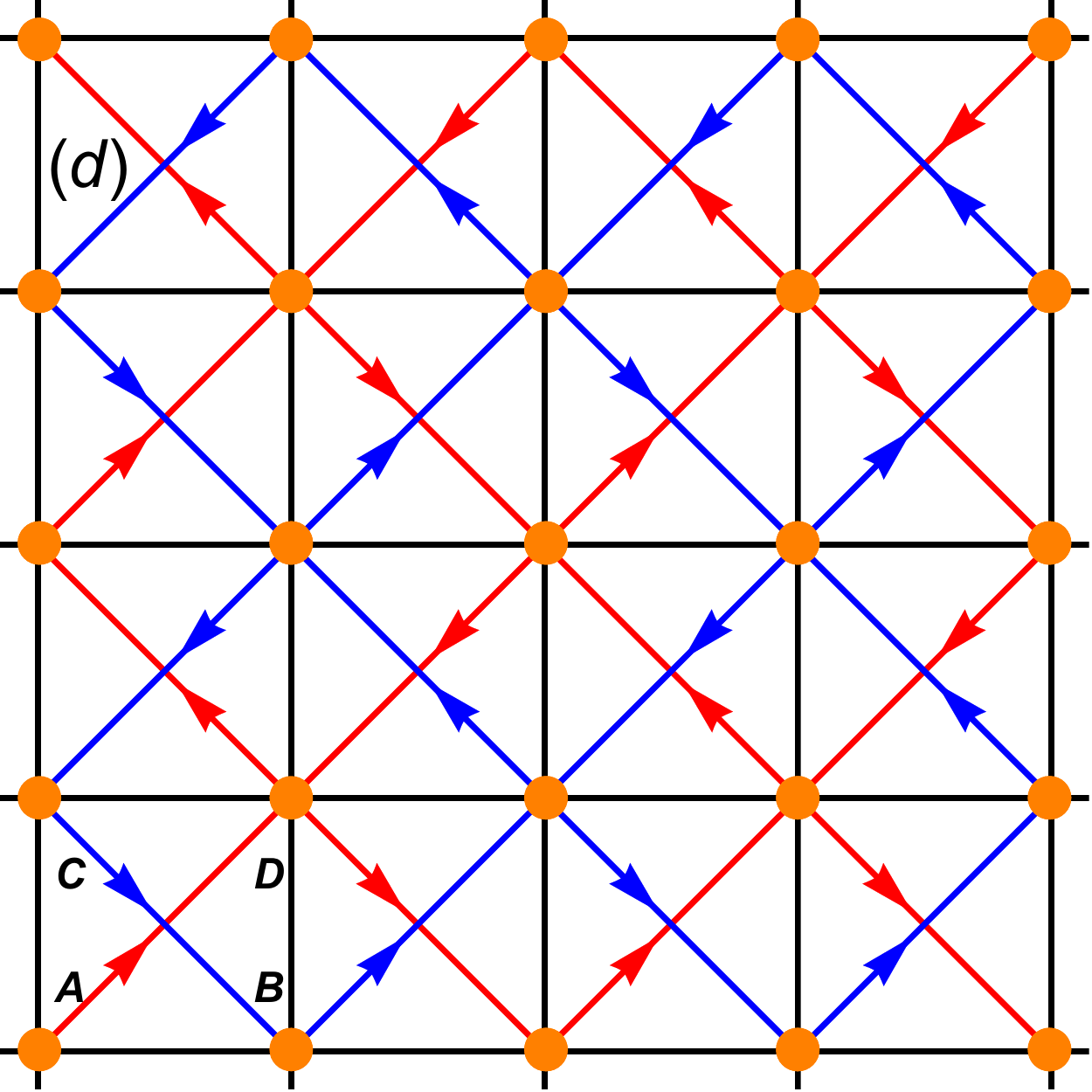}
\caption{(a)~Generalized $\pi$-flux square lattice with real nearest-neighbor (NN) hopping $t_\pm=t(1\pm \beta)$ [suppressed in (b), (c), (d)], supporting linearly dispersing birefringent fermions around ${\bf K}=\left( 1,1\right)\frac{\pi}{2a}$, with $a$ as the lattice spacing. The four-site unit cell is shown by the dashed lines. (b) Quadrupolar charge-density-wave (class CI), where red (blue) dots represent enhanced (depleted) average electronic density. (c) Two patterns (solid and dashed arrows) of translational symmetry breaking current-density-wave (class CII). The imaginary hopping between the NN sites is proportional to $+i$ ($-i$) along the red (blue) arrows. (d) Quantum anomalous Hall insulator (class CIII) resulting from imaginary next-NN hopping.}~\label{Fig:squarelattice}
\end{figure}

Birefringent Dirac fermions can be realized in a \emph{decorated} $\pi$-flux square lattice~\cite{kennett-1, kennett-2, kennett-3, guo-numerics, wang-li-birefringent}, see Fig.~\ref{Fig:squarelattice}(a). Around the Dirac point the quasiparticle spectra $\pm E({\bf k})$ ($+/-$ corresponds to the conduction/valence band), display birefringence with $E({\bf k})= v_\pm |{\bf k}|$. Two effective Fermi velocities are $v_{\pm}=v (1 \pm \beta)$. Similar model can also be realized in graphene~\cite{Dora, Watanabe}, optical lattices~\cite{Lan-1, Lan-2} as well as in three dimensions~\cite{Bradlyn, liangfu, Chen, Boettcher}. For the sake of concreteness, here we solely focus on the planar systems and mostly ignore the real spin degrees of freedom.

\begin{figure}[t!]
\includegraphics[width=0.49\linewidth]{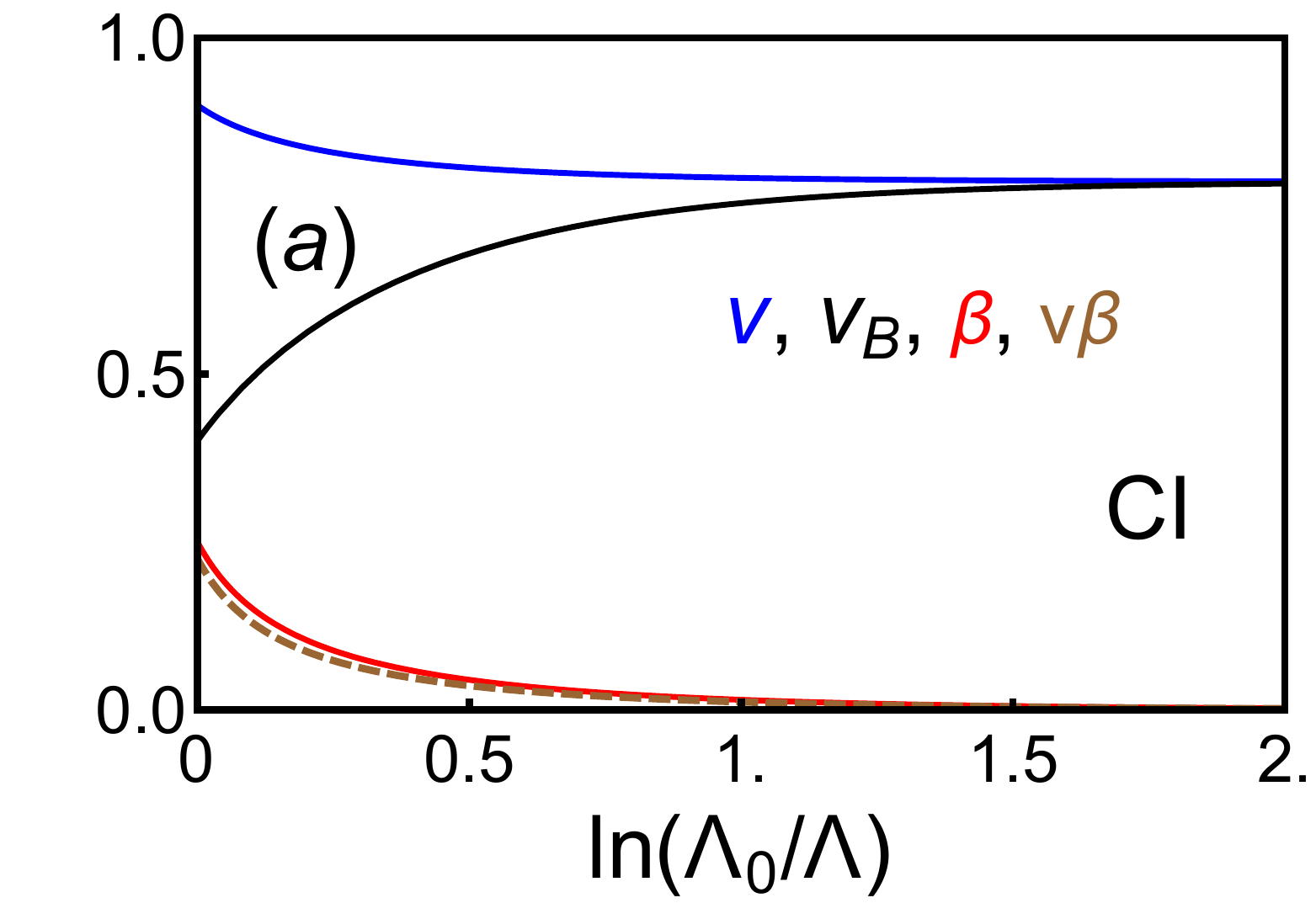}%
\includegraphics[width=0.49\linewidth]{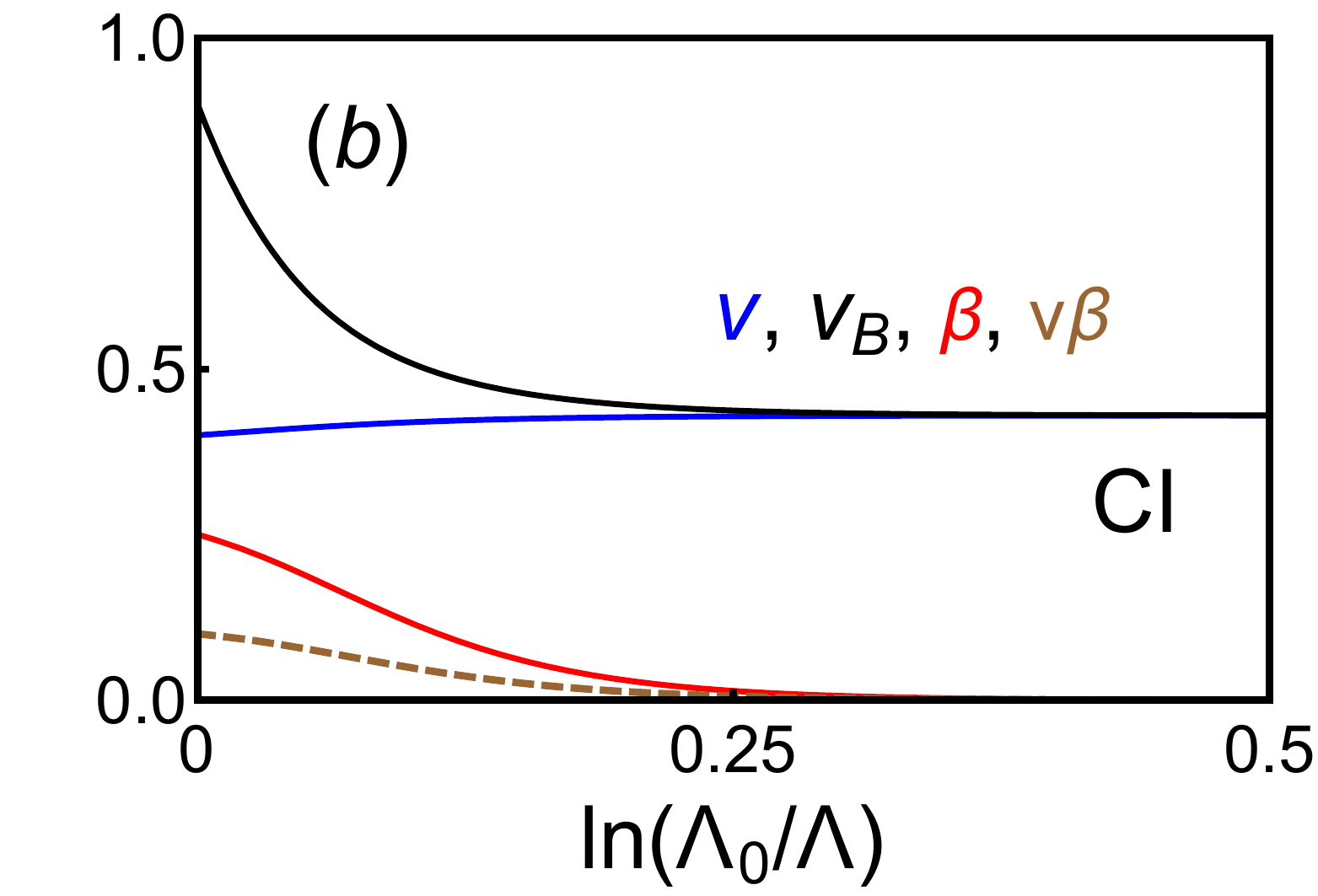}
\includegraphics[width=0.49\linewidth]{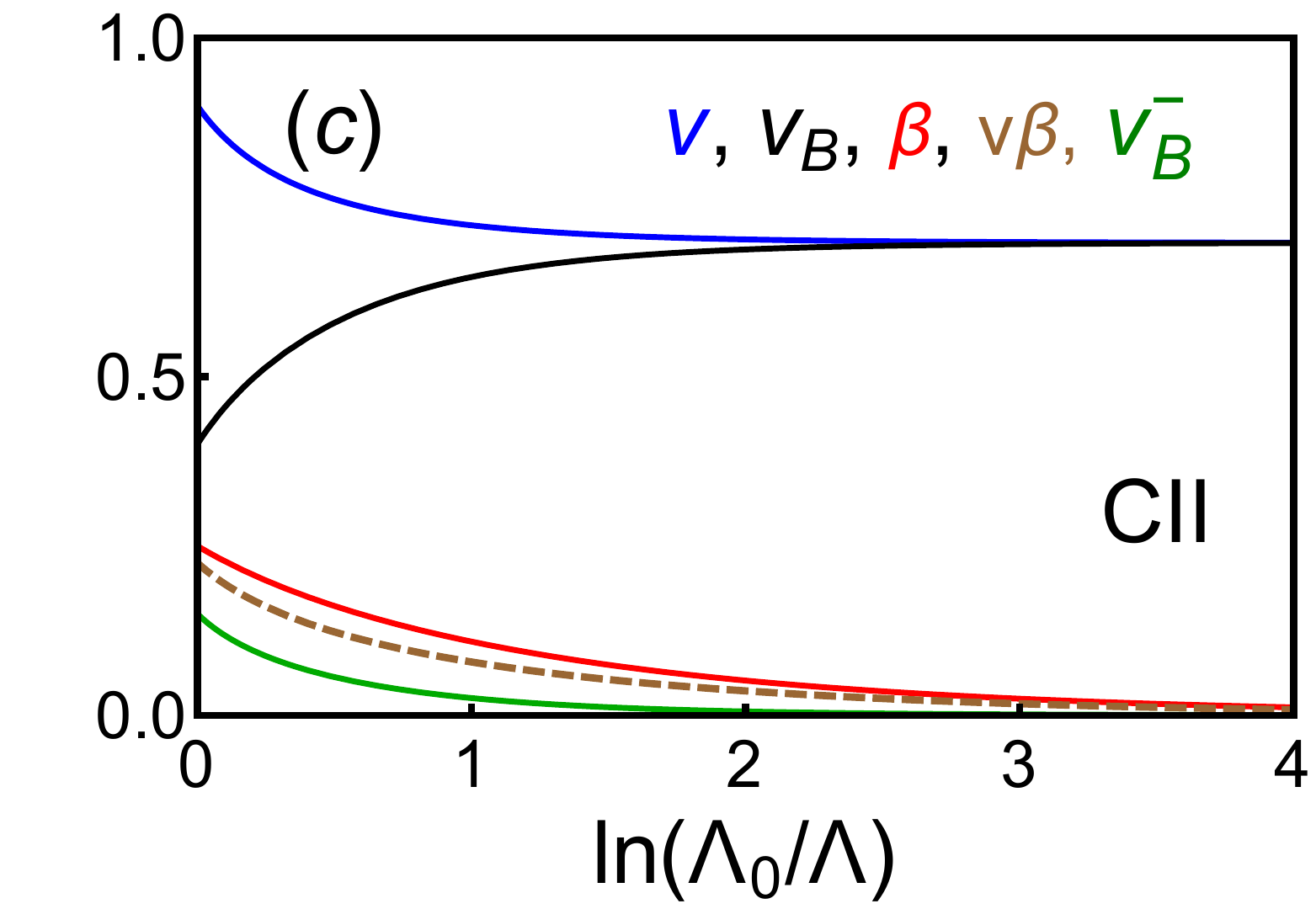}%
\includegraphics[width=0.49\linewidth]{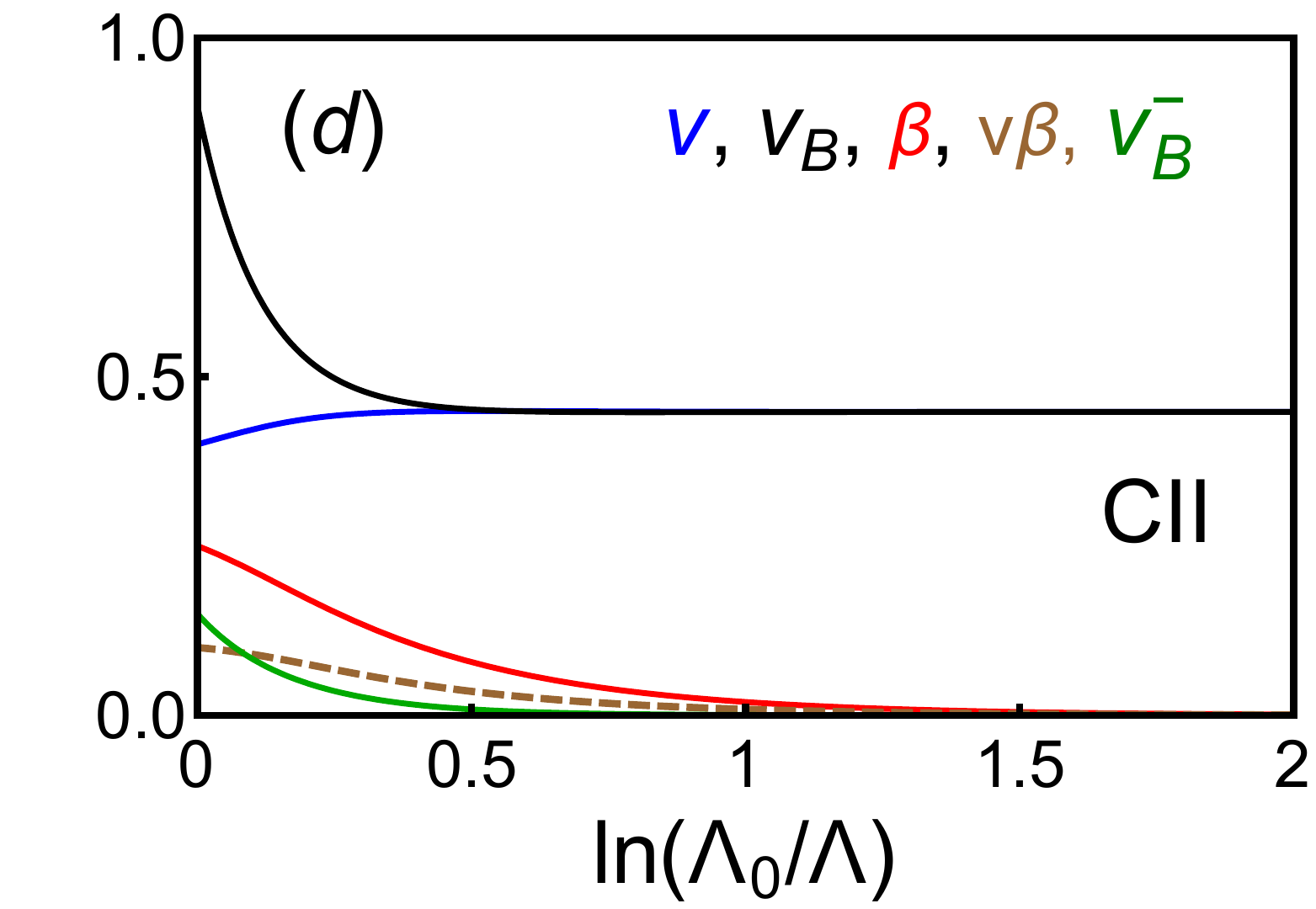}
\includegraphics[width=0.49\linewidth]{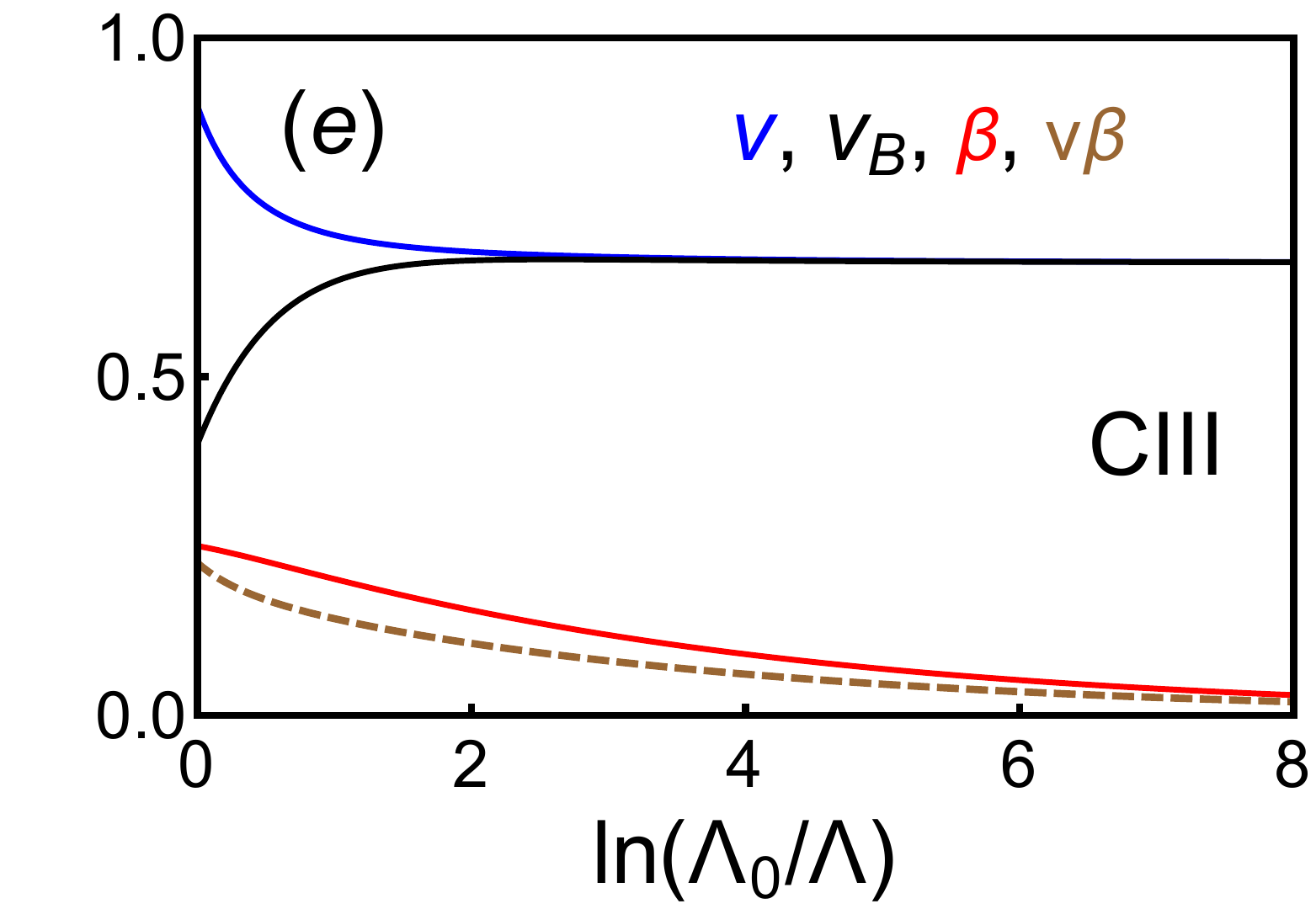}%
\includegraphics[width=0.49\linewidth]{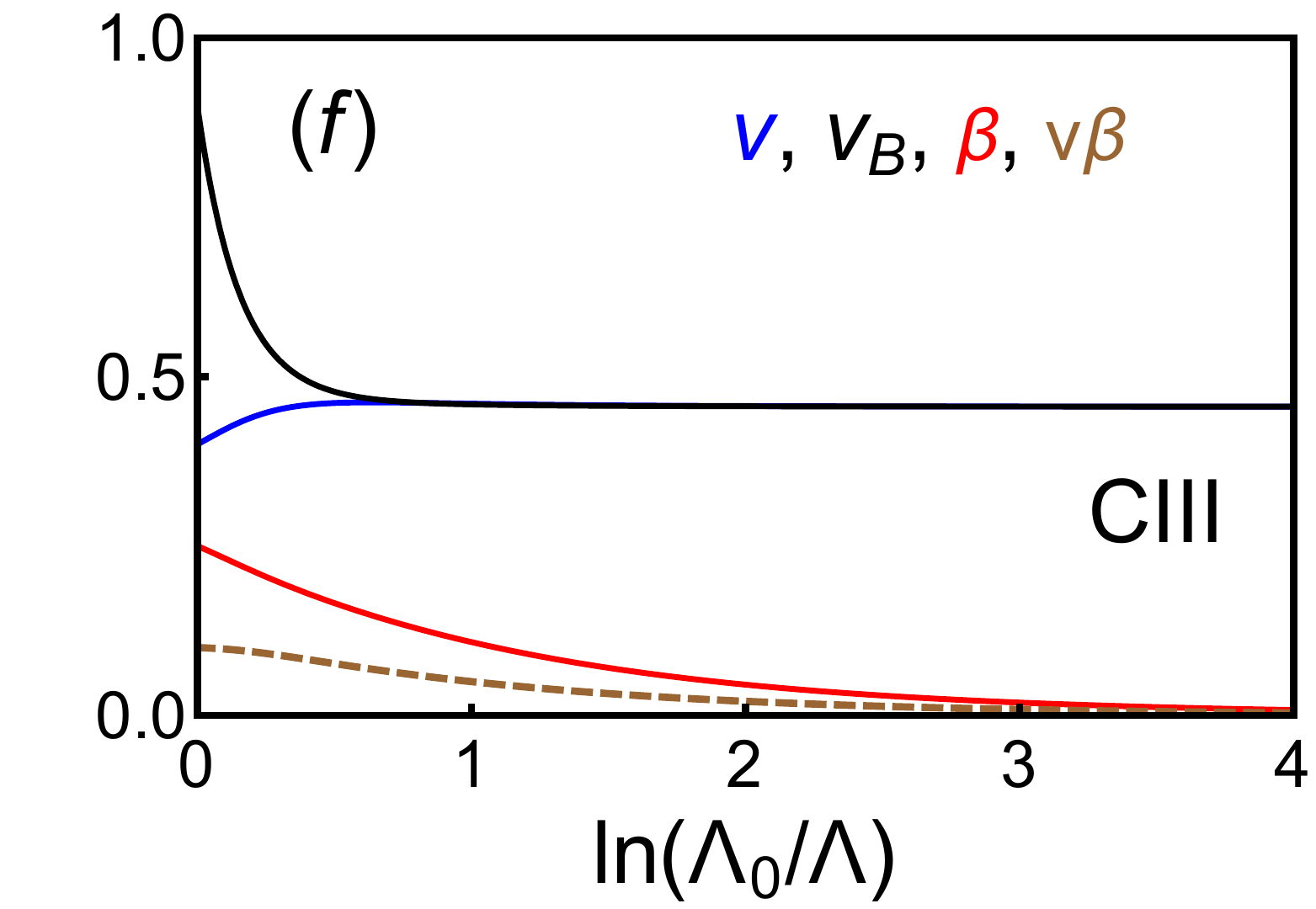}
\caption{RG flow of Fermi velocities $v$ and $v \beta$, bosonic velocities $v_B$ and $v^-_B$ (relevant for only class CII) for $N_f=1$. Irrespective of their bare or initial values, in the deep infrared regime $\ell \equiv \ln(\Lambda_0/\Lambda) \to \infty$, the system possesses a unique terminal velocity near the Mott transition in every class, indicating the restoration of the relativistic or Lorentz symmetry in the vicinity of the Mott-Yukawa QCPs. Here, $\Lambda_0(\Lambda)$ is the bare (running) cutoff. 
}~\label{Fig:RGflowspinless}
\end{figure}

In the absence of birefringence ($\beta=0$), when the local or Hubbardlike interactions among massless Dirac fermions are sufficiently strong, they become susceptible (through continuous QPTs) toward the formation of \emph{four} masses, grouped into two categories: triplet and singlet, respectively transforming as a vector and scalar under the SU(2) chiral rotation~\cite{HJR}, see Table~\ref{table:symmetry}. However, birefringence ($\beta \neq 0$) spoils such an emergent chiral symmetry and four masses fragment into three classes, depending on their (anti)commutation relations with the birefringent component. Nevertheless, we find that the universality classes of the associated continuous QPTs are always determined by that for two copies of spin-1/2 Dirac fermions, and at the QCPs the birefringent velocity disappears ($v \beta \to 0$). Consequently, the system possesses a unique terminal velocity and enjoys an emergent Lorentz symmetry, see Fig.~\ref{Fig:RGflowspinless}. These results also hold for spinful fermions, as well as near the superconducting QCPs.

\emph{Continuum model.}~The Hamiltonian describing a collection of noninteracting pseudospin-3/2 fermions in $d$ spatial dimensions reads $H_{3/2}({\bf k})=H_1^{{\bf k}}+H_2^{{\bf k}}$, where 
\begin{equation}~\label{eq:H32}
H_1^{{\bf k}}=v \Gamma_{j0} k_j, \: H_2^{{\bf k}} = v \beta \; \Gamma_{0j} k_j,
\end{equation}
at low energies~\cite{roy-kennett-yang-juricic}. The momentum ${\bf k}$ is measured from the band touching point, $\Gamma_{\mu\nu}=\sigma_\mu\otimes\sigma_\nu$, where $\{\sigma_\mu\}$ are the Pauli matrices, with $\mu,\nu=0,1,2,3$, while $j=1,2,...,d$. The summation over the repeated indices is assumed. This Hamiltonian features \emph{isotropic} (ensured by the commutation relation $[H_1^{{\bf k}},H_2^{{\bf k}}]=0$) birefringent spectra of linearly dispersing fermionic excitations, and $\beta$ is the birefringence parameter, with $|\beta|<1$.

The birefringent Hamiltonian in Eq.~\eqref{eq:H32} possesses time-reversal symmetry, represented by an antiunitary operator ${\mathcal T}=U_T {\mathcal K}$, with $U_T=\Gamma_{22}$ as its unitary part, ${\mathcal K}$ is the complex conjugation, and ${\mathcal T}^2=+1$ (for spinless fermions). Specifically in two dimensions and in the Dirac limit ($\beta=0$), this Hamiltonian also features an emergent SU(2) chiral symmetry, generated by $\{ \Gamma_{0j} \}$, with $j=1,2,3$, as $[ H^{\bf k}_{1}, \Gamma_{0j}]=0$. It is important to emphasize that the birefringent Hamiltonian $H_{3/2}({\bf k})$ in general does not possess the pseudorelativistic or the chiral SU(2) symmetry. It is only time-reversal symmetric.

\emph{Masses.}~Possible fermion masses $\Psi^\dagger M \Psi$ (with $M$ as a four-dimensional Hermitian matrix) for birefringent fermions and the corresponding pattern of the symmetry breaking (discrete and/or continuous) can be inferred from Eq.~\eqref{eq:H32}. Recall that in the Dirac limit, the system supports four masses $M_\mu=\{\Gamma_{3j}, \Gamma_{30}\}$, with $j=1,2,3$. As $\{ H^{\bf k}_1, M_\mu \}=0$, the quasiparticle spectra inside any mass ordered phase are fully and isotropically gapped. Among them three masses, represented by $\Gamma_{3j}$, form a vector under chiral SU(2) rotations and are time-reversal symmetric. On the other hand, the $\Gamma_{30}$ mass is a chiral scalar, as $[\Gamma_{30},\Gamma_{j0}]=0$, and breaks the ${\mathcal T}$ symmetry. However, the birefringent part of the Hamiltonian $H^{\bf k}_2$ spoils the chiral symmetry and introduces a further fragmentation among them, ultimately yielding the following three symmetry classes for the mass orders, see Table~\ref{table:symmetry}.

(i)~CI: the matrix $M_{\rm CI}=\Gamma_{33}$ anticommutes with $H^{\bf k}_2$,

(ii)~CII: the matrices $M_{{\rm CII},j}=\Gamma_{3j}$, with $j=1,2$, anticommute/commute with one of the matrices in $H^{\bf k}_2$,

(iii)~CIII: the matrix $M_{\rm CIII}=\Gamma_{30}$ commutes with $H^{\bf k}_2$. 
Even when $\beta \neq 0$, the CI and CIII masses produce rotationally symmetric gapped spectra, while such a symmetry is broken in the presence of the CII mass, see Fig.~\ref{Fig:superuniversality}.

\emph{Lattice model.} Two-dimensional birefringent fermions can be realized in a decorated $\pi$-flux square lattice with real nearest-neighbor (NN) hopping amplitudes~\cite{kennett-1, kennett-2, kennett-3, guo-numerics, wang-li-birefringent}, as shown in Fig.~\ref{Fig:squarelattice}(a). The corresponding lattice Hamiltonian is $H_0=\sum_{\bf p}\Psi_{\bf p}^\dagger H_0({\bf p})\Psi_{\bf p}$, where
\allowdisplaybreaks[4]
\begin{equation}~\label{eq:tightbinding}
H_0({\bf p})=2\sum_{mn}t_m \left[ \delta_{mn} \Gamma_{n1} \cos (p_x a)+ \epsilon_{mn} \Gamma_{1n} \cos (p_y a) \right],    
\end{equation}
the indices $m,n=+,-$, $\delta_{mn}$ ($\epsilon_{mn}$) is the Kronecker delta (antisymmetric Levi-Civita symbol, with $\epsilon_{+-}=1$), $t_\pm=t(1\pm\beta)$, $\sigma_\pm=(\sigma_0\pm\sigma_3)/2$, and ${\bf p}$ is the lattice momentum. The components of the spinor annihilation operator $\Psi_{\bf p}^\top\equiv(A_{\bf p}, B_{\bf p}, C_{\bf p}, D_{\bf p})$ act on the Hilbert space of the states, localized on the corresponding site of the $\pi$-flux lattice. Expanding $H_0({\bf p})$ about the band touching point at ${\bf K}=(1,1)\frac{\pi}{2a}$ and performing a unitary rotation by 
\begin{equation}~\label{eq:Unitaryrotation}
U=\frac{1}{\sqrt{2}} \left[ \Gamma_{+0} +i\Gamma_{-3}
+\frac{1}{2} \left( \Gamma_{11} + \Gamma_{12} +i \Gamma_{21}-i\Gamma_{22} \right)\right],
\end{equation}
under which $\Psi_{\bf p} \rightarrow U^\dagger \Psi_{\bf p}$, we obtain $H_{3/2}({\bf k})$, see Eq.~\eqref{eq:H32}, with ${\bf k}={\bf K}-{\bf p}$ and $v=ta$. Next we discuss lattice realizations of various mass orders for birefringent fermions.

A charge-density-wave of \emph{quadrupolar} arrangement within the four-site unit cell is shown in Fig.~\ref{Fig:squarelattice}(b) and represented by the matrix $\Gamma_{33}$ in the announced spinor basis. This matrix retains its form after the unitary transformation by $U$, and fully anticommutes with $H^{\bf k}_{2}$. Hence, such a charge-density-wave order belongs to class CI for birefringent spin-$3/2$ fermions, see Table~\ref{table:symmetry}.

The order parameters belonging to class CII break the translational symmetry in a square lattice and represent \emph{current-density-waves}. Two possible realizations of such an order are shown in Fig.~\ref{Fig:squarelattice}(c). The corresponding fermionic bilinears appear with the matrices $\Gamma_{02}$ (solid arrows) and $\Gamma_{23}$ (dashed arrows)~\cite{TRS-comment}. After the unitary rotation by $U$ they respectively transform into $M_{{\rm CII},j}=\Gamma_{3j}$ for $j=1$ and $2$. They neither fully commute nor fully anticommute with $H^{\bf k}_2$, see Table~\ref{table:symmetry}.

The order parameter from class CIII fully commutes with $H^{\bf k}_2$ and is represented by purely \emph{imaginary} next-NN hopping between the pairs of sites $A$ and $D$, and $B$ and $C$ of the square lattice, as shown in Fig.~\ref{Fig:squarelattice}(d). Such an ordered phase breaks the time-reversal symmetry and represents a quantum anomalous Hall insulator~\cite{kennett-3, wang-li-birefringent}. The corresponding matrix $\Gamma_{12}$, appearing in the order-parameter, after the unitary rotation by $U$, transforms into $M_{\rm CIII}=U^\dagger \Gamma_{12} U=\Gamma_{30}$, see Table~\ref{table:symmetry}.

\begin{table}[t!]
\centering
\begin{tabular}{ |c|c|c|c|c|c| } 
\hline
Class & Mass matrix (${\bf M}$) & Chiral SU(2) & TRS & $\Gamma_{01}$ & $\Gamma_{02}$ \\
\hline \hline
CI & $\Gamma_{33}$ & Vector & \cmark & $-$ & $-$ \\
\hline
\multirow{2}{1.5em}{CII} & $\Gamma_{31}$ & Vector & \cmark & $+$ & $-$ \\ & $\Gamma_{32}$ & Vector & \cmark & $-$ & $+$ \\ 
\hline
CIII & $\Gamma_{30}$ & Scalar & \xmark & $+$ & $+$ \\
\hline
\end{tabular}
\caption{Order parameters in three classes (first column), together with their matrix representations (second column). The third column shows their symmetry transformation under the SU(2) chiral rotation, generated by $\{ \Gamma_{01}, \Gamma_{02}, \Gamma_{03} \}$. The fourth column shows whether the order parameter is time-reversal symmetric (\cmark) or not (\xmark). The last two columns show if the order parameter commutes ($+$) or anticommutes ($-$) with the matrices $\Gamma_{01}$ and $\Gamma_{02}$, appearing
in the birefringent part of the noninteracting Hamiltonian $H^{\bf k}_2$, see Eq.~\eqref{eq:H32}. }
\label{table:symmetry}
\end{table}

\emph{RG analysis.}~After establishing a one-to-one correspondence between the continuum description of spin-3/2 birefringent fermions and their lattice realization, we perform a renormalization group (RG) analysis of the quantum-critical theory near the above three classes of mass orderings. To show the robustness of the emergent superuniversality, we mainly focus on the RG flow equations for the velocities of birefringent fermions and bosonic order parameter fluctuations ($v_B$) in the quantum-critical region. The corresponding space-(imaginary) time ($\tau$) action is $S=S_{\rm F}+S_{\rm BF}+S_{\rm B}$, where
\begin{equation}
S_{\rm F}=\int \Psi^\dagger(\tau,{\bf r})\left[\partial_\tau + H_{3/2}({\bf k}\rightarrow -i\nabla)\right]\Psi(\tau,{\bf r}),
\end{equation}
and $\int \equiv \int d\tau d^d{\bf r}$.~The elements of the four-component spinor $\Psi(\tau,{\bf r})$ depend on microscopic details.~The fermionic Matsubara frequency Green's function reads
\allowdisplaybreaks[4]
\begin{align}~\label{eq:fermion-propagator}
G^{\bf k}_{\rm F}(i\omega)=\left[ i\omega+ H_{3/2}({\bf k}) \right] \frac{\left[\omega^2+ v^2_{\rm av} {\bf k}^2-2H_1^{{\bf k}}H_2^{{\bf k}}\right]}{(\omega^2+v_+^2 {\bf k}^2)(\omega^2+v_-^2 {\bf k}^2)},\nonumber
\end{align}
where $v^2_{\rm av}= (v^2_+ + v^2_-)/2$. The action for the real bosonic order-parameter fields $\Phi_\alpha\equiv\Phi_\alpha(\tau,{\bf r})$ has the form
\begin{align}
S_{\rm B} &=\sum^{N_b}_{\alpha=1} \int \left\{ \frac{1}{2}\left[(\partial_\tau\Phi_\alpha)^2 + v_B^2(\partial_j\Phi_\alpha)^2+m^2_B \Phi_\alpha^2 \right]\right. \nonumber\\
&\left.+\frac{\lambda_0}{4!}(\Phi_\alpha^2)^2 \right\}
\equiv S_{\rm B}^0+S_{\rm B}^{\rm int},
\end{align}
where $N_b=1 (2)$ for CI and CIII (CII), $m^2_B$ is the tuning parameter for the transition, and $\lambda_0$ is the bare coupling constant for the $\Phi^4$ interaction. The Green's function for free bosonic field in the critical hyperplane ($m^2_B=0$) is
\begin{equation}
G^{\bf k}_{{\rm B},\alpha\beta}(i\omega)
\equiv G^{\bf k}_{\rm B}(i\omega)\delta_{\alpha\beta}=\frac{1}{\omega^2+v_B^2 {\bf k}^2}
\,\,\delta_{\alpha\beta}. \nonumber 
\end{equation}
The Yukawa coupling between gapless birefringent excitations and bosonic order-parameter field takes the form
\begin{equation}
S_{\rm BF}=g_0 \; \sum^{N_b}_{\alpha=1}\int \,\Phi_{\alpha}\,\Psi^\dagger (\tau,{\bf r}) M_\alpha \Psi(\tau,{\bf r}),
\end{equation}
as the latter is a composite object of two fermionic fields. We compute the RG flow equations by integrating out the fast modes with Matsubara frequency $-\infty<\omega<\infty$ and within the Wilsonian momentum shell $\Lambda {\rm e}^{-\ell}<|{\bf k}|<\Lambda $. Here $\ell(>0)$ is the logarithm of the RG scale. The momentum integrals are performed around $d=3$, the upper critical (spatial) dimension of the theory~\cite{zinn-justin}, while the matrix algebra is carried out in $d=2$~\cite{roy-goswami-jusiric}.

We now present the RG flow analysis for the velocities for each of the three classes. They are obtained from the corresponding leading order fermionic and bosonic self-energy diagrams at external Matsubara frequency ($i\nu$) and momentum (${\bf k}$)~\cite{roy-juricic-herbut-jhep}, respectively given by 
\begin{eqnarray}
\Sigma^{\bf k}_{\rm F}(i\nu)=g^2_0 \sum^{N_b}_{\alpha=1} \int_{\omega,{\bf q}} M_\alpha G^{\bf q}_{\rm F} (i\omega) M_\alpha 
G^{{\bf k}-{\bf q}}_{\rm B}(i\omega-i\nu),
\end{eqnarray}
where $\int_{\omega,{\bf q}} \equiv \int d\omega d^d{\bf q}/(2\pi)^{d+1}$, and
\begin{eqnarray}
\Pi^{\bf k}_{\rm B}(i\nu)=-\frac{1}{2} \; g^2_0 \; {\rm Tr} \int_{\omega,{\bf q}} M_\alpha G^{\bf q}_{\rm F}(i\omega) M_\alpha G^{{\bf k}+{\bf q}}_{\rm F}(i\omega+i\nu).
\end{eqnarray}
The trace is taken over the Dirac matrices $\Gamma_{\mu\nu}$. We then use the standard renormalization conditions to first obtain the renormalization factors for the fermionic and bosonic fields that in turn yield the flow equations for the fermionic and bosonic  velocities, as well as for the birefringent parameter, analogously as in Ref.~\cite{roy-kennett-yang-juricic}.

\emph{Class CI}. The quadrupolar charge-density wave ordering has already been addressed in Ref.~\cite{roy-kennett-yang-juricic}, which we present here for the sake of completeness. The RG flow equations for the velocities read as
\allowdisplaybreaks[4]
\begin{align}
\frac{dv}{d\ell} &= - 2 N_b g^2 v \; A, 
\frac{d \left(v\beta \right)}{d\ell}= -2 N_b g^2 \left( v \beta \right) \; Q_+, \\
\frac{d v_B}{d \ell} &= -\frac{g^2 N_f \; v_B}{2 v^3(1-\beta^2)} \left[ \frac{(1+\beta^2)^2}{(1-\beta^2)^2}
- \frac{v^2}{v^2_B} \left\{ 1+\frac{2}{3} \beta^2 \right\} \right], \nonumber 
\end{align}
where $N_f$ is the flavor number for spin-3/2 fermions, $N_b=1$ is the number of bosonic order parameter components for an Isinglike orderparameter, $g^2 = g^2_0 k^{-\epsilon}/(8\pi^2)$ is the dimensionless Yukawa coupling, $\epsilon=3-d$, $Q_\zeta= B + \zeta M$ with $\zeta=+$, $B_1=(v+v_B)^2$, $B_2=\beta^2 v^2$, and
\begin{align}
A &= \frac{2(v-v_B)B_1+4 v B_2}{3 v v_B(B_1-B_2)^2},
B = \frac{B_1+B_2}{v_B(B_1-B_2)^2},\nonumber\\
M &= \frac{2B_1-B_2+v_B^2-v^2}{3v_B (B_1-B_2)^2}.
\end{align}
Irrespective of the relative magnitude of bare Fermi and boson velocities, we always find that they reach a common terminal velocity and $v \beta \to 0$ in the deep infrared regime ($\ell \to \infty$), as shown in Figs.~\ref{Fig:RGflowspinless}(a) and~\ref{Fig:RGflowspinless}(b).

\emph{Class CII}. In this case, the Yukawa interaction generates an anisotropy between the bosonic velocities in two directions. The RG flow equations (with $N_b=2$) are
\begin{align}
\frac{d v}{d \ell} &= -2 N_b g^2 v A, \:\:
\frac{d (v \beta)}{d \ell}= -2 N_b g^2 Q_0, \nonumber \\
\frac{d v^\tau_{_B}}{d \ell} &=- \frac{g^2 N_f}{2 v^3} v^\tau_{_B} \; \left[ C(\beta)
- \tau \frac{v^2 \; Y_{-\tau} (v^\tau_{_B}, v^{-\tau}_{_B},\beta)}{(v^+_{_B} + v^-_{_B})(1-\beta^2)} \right], 
\end{align}
for $\tau=\pm$, where $v^\tau_{_B}=(v^x_{_B}+\tau v^y_{_B})/2$, $v^+_{_B} \equiv v_B$ and
\begin{align}
C(\beta) &= \frac{1}{(1-\beta^2)^3} \left( 1 + \frac{3}{2} \beta^2 -\frac{1}{2} \beta^6 \right), \nonumber \\
Y_{\pm} (a,b,\beta) &= \frac{1}{|a-b|} \left( 1+ \frac{2\beta^4}{15} \right) \pm \frac{\beta^2}{3 a} \left( 1-\frac{\beta^2}{5} \right).
\end{align} 
Note that $dv^-_{_B}/d\ell$ is a negative definite quantity for $|\beta|<1$. Hence, the anisotropy in the boson velocity is an \emph{irrelevant} quantity, which otherwise solely arises for finite $\beta$. In the deep infrared regime $v \beta, v^-_B \to 0$, see Figs.\ref{Fig:RGflowspinless}(c) and~\ref{Fig:RGflowspinless}(d), and we enjoy the liberty of computing the fermionic self-energy by setting $v^-_B=0$ from the outset. Also, after a long RG time the average Fermi and boson velocities approach a common terminal velocity.

\emph{Class CIII}. The leading order RG flow equations in the close proximity of the class CIII mass ordering read
\begin{align}
\frac{dv}{d\ell}      &= - 2 N_b v g^2 \; A, \:
\frac{d (v\beta)}{d\ell}  = - 2 N_b (v\beta) g^2 \; Q_-, \nonumber \\
\frac{dv_{_B}}{d\ell} &= - \frac{g^2 N_f \; v_{_B}}{2 v^3} \left[ 1-\frac{v^2}{v^2_{_B}} \left\{ \frac{3-6 \beta^2+\beta^4}{3(1-\beta^2)} \right\} \right],
\end{align}
with $N_b=1$. Once again we find that in the deep infrared regime $v \beta \to 0$, and boson and Fermi velocity acquire a common terminal velocity, see Figs.~\ref{Fig:RGflowspinless}(e) and~\ref{Fig:RGflowspinless}(f).

In all three cases we find that when a collection of strongly interacting spin-3/2 fermions resides at the brink of spin-singlet Mott insulation, the system possesses a \emph{unique} velocity, indicating a generic and robust restoration of the Lorentz symmetry near the Mott-Yukawa QCPs. Moreover, as $v \beta \to 0$ at this QCP, the associated quantum critical regime supports two copies of spin-1/2 critical Dirac fermions, strongly coupled with fluctuating bosonic orderparameter excitations (without any long range ordering), constituting a relativistic non-Fermi liquid, devoid of any sharp quasiparticle excitations. We also note that the requisite RG time ($\ell_\star$) for the restoration of the Lorentz symmetry decreases as the number matrices from the birefringent Hamiltonian $H^{\bf k}_2$, anticommuting with the mass matrix increases, and $\ell^{\rm CI}_\star < \ell^{\rm CII}_\star < \ell^{\rm CIII}_\star$. See Fig.~\ref{Fig:RGflowspinless} and compare with Table~\ref{table:symmetry}.

On the Lorentz symmetric critical hyperplane, defined by $v=v{_B}=1,\beta=0, m^2_B=0$, the rest of the RG flow equations simplify to    
\begin{equation}
\frac{d g^2}{d\ell}=g^2 \left[\epsilon - b_1 g^2\right], \: 
\frac{d\lambda}{d\ell}= \epsilon \lambda - 4 N_f g^2 \left[ \lambda-6g^2\right]-\frac{b_2 \lambda^2}{6},
\end{equation}
where $\lambda=\lambda_0 k^{-\epsilon}/(8\pi^2)$ is the dimensionless bosonic self-interaction coupling, $b_1=2 N_f+4-N_b$, and $b_2=N_b+8$. The Mott-Yukawa QCP is located at 
\begin{equation}
(g^2_\ast,\lambda_\ast)=\left(1, \frac{3}{b_2} [b_3+ \sqrt{b^2_3+ 16 N_f b_2 }] \right) \frac{\epsilon}{b_1},
\end{equation}  
where $b_3=4-2N_f-N_b$. The bosonic and fermionic anomalous dimensions at this QCP are $\eta_\Phi=2 N_f g^2_\ast$ and $\eta_\Psi=N_b g^2_\ast/2$, respectively, and the residue of the fermionic quasiparticle pole vanishes as $Z_\Psi \sim (m_F)^{\eta_\Psi/2}$. The fermionic mass ($m_F$) vanishes according to a universal ratio $m^2_B/m^2_F \sim \lambda_\ast/g_\ast$, as the QCP is approached from the ordered side. The correlation length exponent (obtained from the RG flow of $m^2_B$, not shown explicitly) for the QPT is $\nu=1/2 + N_f g^2_\ast/2+ (N_b+2)\lambda_\ast/24$.

We note that during the course of the RG transformation in the class CI and CIII, new fermion bilinears 
\begin{equation}
(v \beta) \frac{2 N_b (v+v_{_B}) g^2}{v_{_B} \left[ (v+v_{_B})^2 -v^2 \beta^2 \right]^2} \left( \sum^{d}_{j=1} \Psi^\dagger \left(i  \Gamma_{0j} \Gamma_{j0} \right) \Psi \right) \frac{1}{\epsilon}\nonumber 
\end{equation}
get generated. But, the RG transformation is performed in such a way that the generated terms can be eliminated by the corresponding \emph{counterterms} and $H_{3/2}({\bf k})$ retains its original form under the coarse graining. This procedure can also be justified \emph{a posteriori} by noticing that the generated terms are proportional to the birefringent velocity $v \beta$, which ultimately vanishes at all the QCPs, see Fig.~\ref{Fig:RGflowspinless}. No new term gets generated (at least to the leading order) in class CII.

\emph{Discussion.}~To summarize, we here show that strongly interacting birefringent fermions, residing sufficiently close to itinerant Mott QCPs, possess a superuniversal description: \emph{relativistic non-Fermi liquid}, irrespective of the nature of the mass ordering. We expect these conclusions to hold (at least qualitatively) even when we include the full gauge interactions, with the only difference that the common terminal velocity is the velocity of light ($c$), as shown for spin-1/2 Dirac systems~\cite{roy-juricic-herbut-jhep}.

With the restoration of spin degrees of freedom, each of mass can now be realized in spin-singlet (discussed so far) and spin-triplet channels. In a decorated $\pi$-flux square lattice, spin-triplet class CI (CIII) order corresponds to quadrupolar spin-density-wave~\cite{roy-kennett-yang-juricic} (quantum-spin Hall insulator), for which the above discussion can be generalized by setting $N_b=3$. By contrast, class CII order corresponds to spin-triplet current-density-wave. Critical behavior near such an ordering can be addressed by setting $N_b=2=4-2$, where 4 (2) is the number of mutually anti-commuting (commuting) mass matrices~\cite{roy-juricic-SCMCP}. Therefore, our conclusions regarding the generic restoration of Lorentz symmetry and emergence of robust superuniversality are equally applicable for spin-triplet Dirac insulators. These predictions can at least be tested from quantum Monte Carlo simulations of Hubbard~\cite{guo-numerics, fakher-piflux,sorella,thomaslang-PRL} and extended-Hubbard models, for example.

When the finite range interactions acquire sufficiently strong attractive components, birefringent fermions can develop strong propensities toward the nucleation of a plethora of superconducting phases. Mundane spin-singlet $s$-wave pairing fully anticommutes with $H^{\bf k}_2$, and belongs to class CI. Remaining three spin-triplet mass pairing for regular Dirac fermions fragment into two classes. Namely, two translational symmetry breaking pairings or pair-density-waves (analogous to two Kekul\'{e} superconductors in honeycomb lattice~\cite{roy-herbut-kekule}) belong to class CII, and the remaining triplet $p$-wave pairing (analogous to the $f$-wave pairing in honeycomb lattice~\cite{honerkamp}) belongs to class CIII. While the restoration of the Lorentz symmetry and emergence of superuniversality for class CI $s$-wave pairing have already been demonstrated in Ref.~\cite{roy-kennett-yang-juricic}, based on the discussion presented here we expect that these outcomes should also hold for class CII and class CIII pairings, which we demonstrate explicitly in future.

\emph{Acknowledgments}. B.R. was supported by the Startup Grant from Lehigh University.



\end{document}